\documentclass[letterpaper]{article} 
\usepackage{aaai2026}  
\usepackage{times}  
\usepackage{helvet}  
\usepackage{courier}  
\usepackage[hyphens]{url}  
\usepackage{graphicx} 
\usepackage{amsmath}
\usepackage{natbib}  
\usepackage{caption} 
\usepackage{algorithm}
\usepackage{algorithmic}
\usepackage{tabularx}
\usepackage{booktabs}
\usepackage{graphicx}
\usepackage{subcaption}
\usepackage{amssymb}

\usepackage{newfloat}
\usepackage{listings}
\DeclareCaptionStyle{ruled}{labelfont=normalfont,labelsep=colon,strut=off} 
\floatstyle{ruled}
\newfloat{listing}{tb}{lst}{}
\floatname{listing}{Listing}
\title{Regressor-guided Diffusion Model \\ for De Novo Peptide Sequencing with Explicit Mass Control}
\author{
    Shaorong Chen\textsuperscript{\rm 1,2}, Jingbo Zhou\textsuperscript{\rm 1,2}, Jun Xia\textsuperscript{\rm 3,4 }$^{\dagger}$
}
\affiliations{
    \textsuperscript{\rm 1} Zhejiang University, Hangzhou, China, 310058\\
    \textsuperscript{\rm 2} AI Lab, Westlake University, Hangzhou, China, 310030\\
    \textsuperscript{\rm 3} AIMS Lab, The Hong Kong University of Science and Technology (Guangzhou), Guangzhou, China, 511453\\
    \textsuperscript{\rm 4} The Hong Kong University of Science and Technology, Hong Kong, China, 999077\\

    chenshaorong@westlake.edu.cn, junxia@hkust-gz.edu.cn
}

\usepackage{bibentry}

\begin{document}

\maketitle

\begin{abstract}
The discovery of novel proteins relies on sensitive protein identification, for which de novo peptide sequencing (DNPS) from mass spectra is a crucial approach. While deep learning has advanced DNPS, existing models inadequately enforce the fundamental mass consistency constraint—that a predicted peptide's mass must match the experimental measured precursor mass. Previous DNPS methods often treat this critical information as a simple input feature or use it in post-processing, leading to numerous implausible predictions that do not adhere to this fundamental physical property. To address this limitation, we introduce DiffuNovo, a novel regressor-guided diffusion model for de novo peptide sequencing that provides explicit peptide-level mass control. Our approach integrates the mass constraint at two critical stages: during training, a novel peptide-level mass loss guides model optimization, while at inference, regressor-based guidance from gradient-based updates in the latent space steers the generation to compel the predicted peptide adheres to the mass constraint. Comprehensive evaluations on established benchmarks demonstrate that DiffuNovo surpasses state-of-the-art methods in DNPS accuracy. Additionally, as the first DNPS model to employ a diffusion model as its core backbone, DiffuNovo leverages the powerful controllability of diffusion architecture and achieves a significant reduction in mass error, thereby producing much more physically plausible peptides. These innovations represent a substantial advancement toward robust and broadly applicable DNPS. The source code is available in the supplementary material.

\end{abstract}


\section{Introduction}

\begin{figure}[t]
    \centering
    \includegraphics[width=1\linewidth]{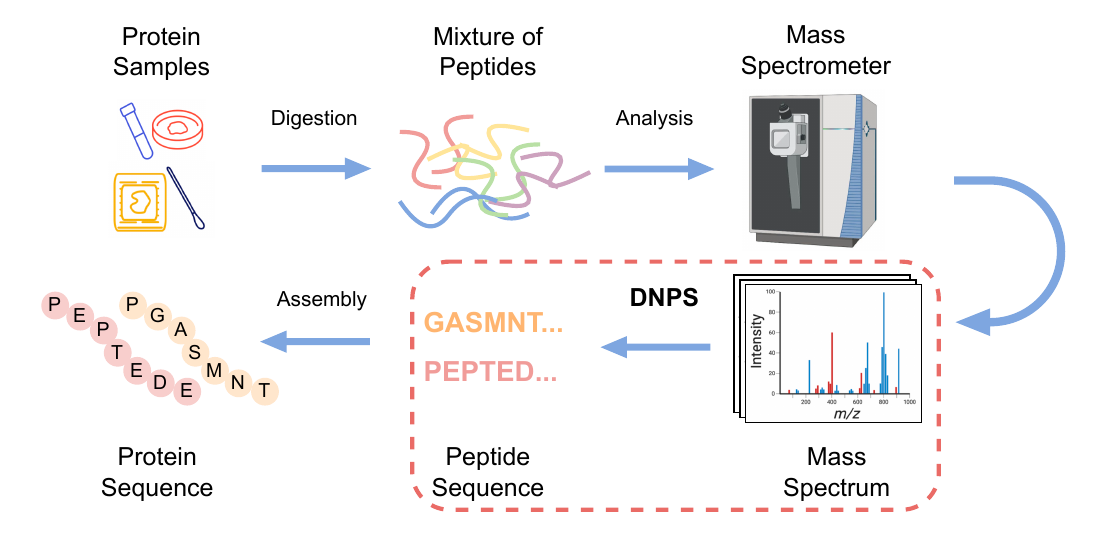}
    \caption{Schematic of a typical bottom-up proteomics workflow \cite{zhang2013protein}. Proteins are digested into peptides and analyzed by mass spectrometer to produce mass spectrum. DNPS is the process of inferring a peptide's sequence directly from its spectrum. The resulting peptide sequences can then be used for protein sequence assembly.}
    \label{fig:dnps}
\end{figure}
\begin{figure}[h] 
    \centering 

    \begin{subfigure}[b]{0.48\columnwidth}
        \centering
        \includegraphics[width=\textwidth]{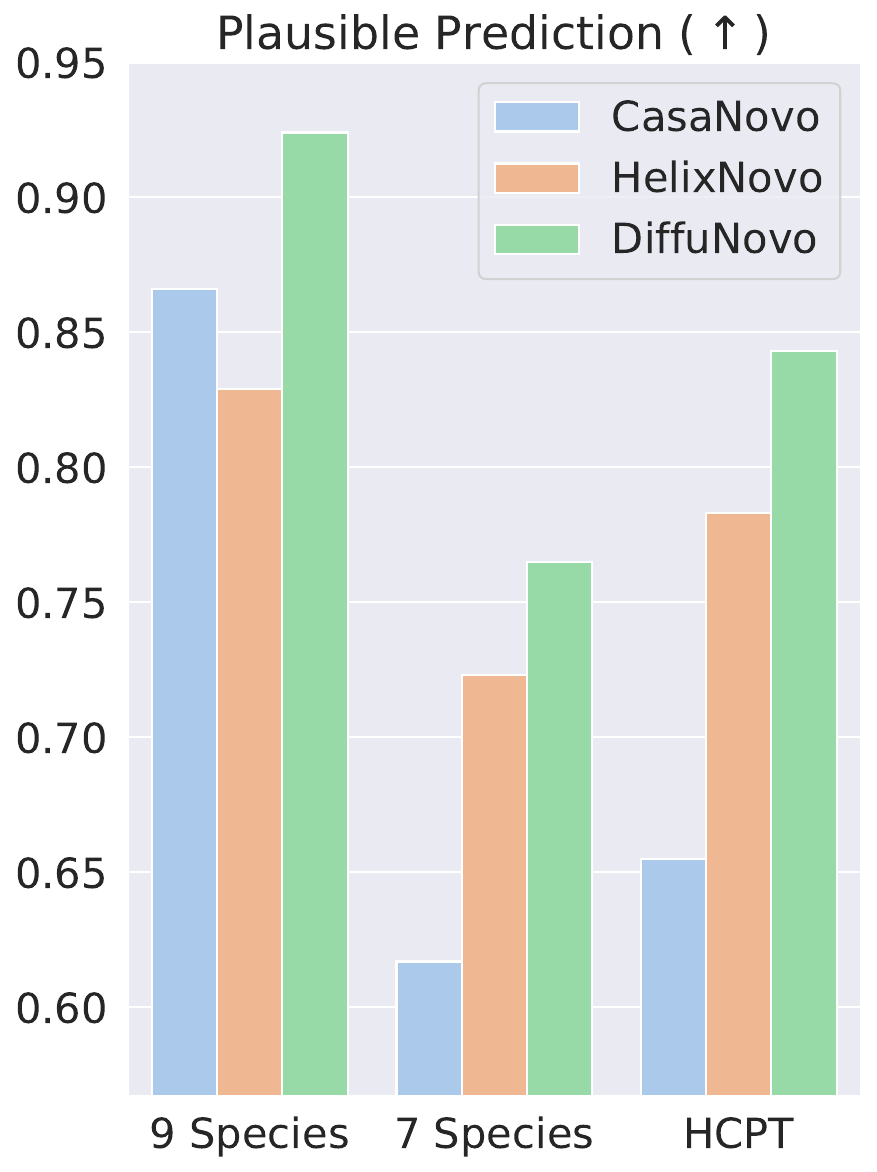}
        \subcaption{Plausible Prediction Rate ($\uparrow$)} 
        \label{fig:sub_a} 
    \end{subfigure}
    \begin{subfigure}[b]{0.48\columnwidth}
        \centering
        \includegraphics[width=\textwidth]{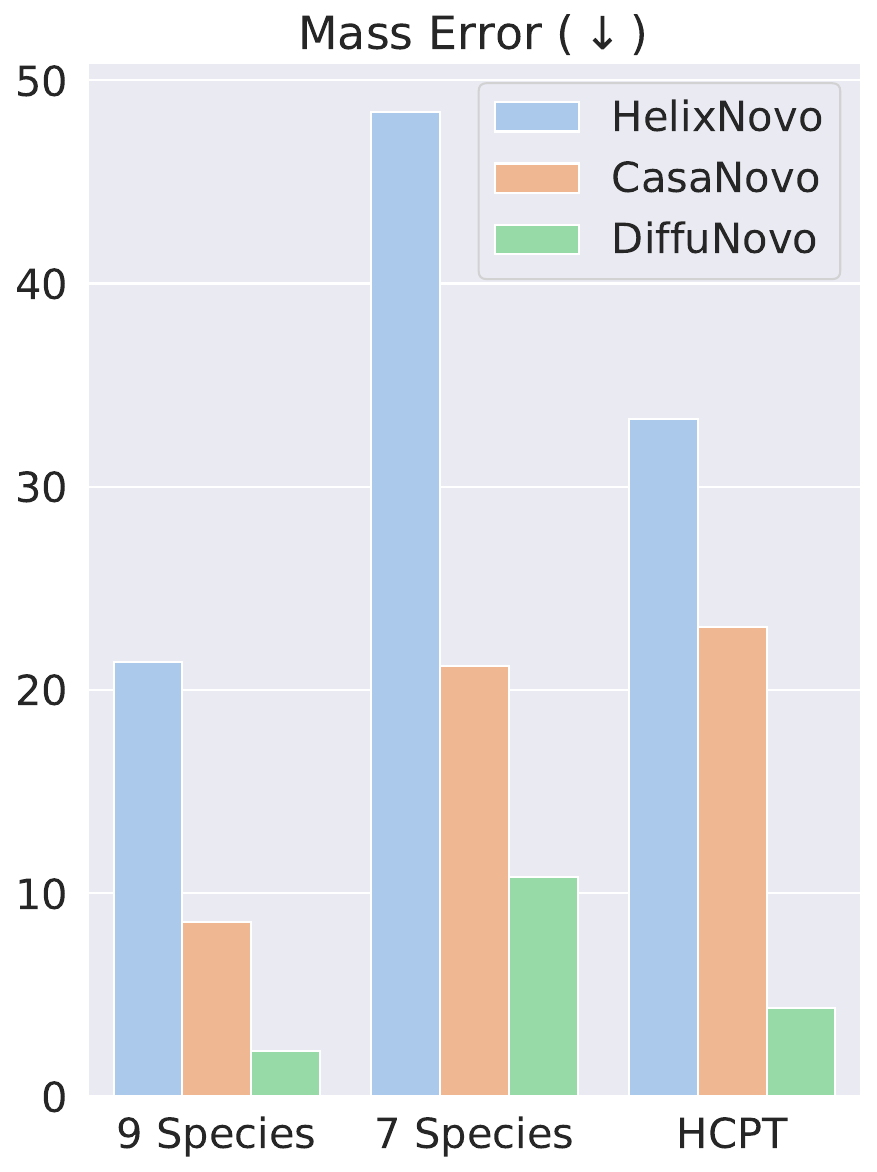}
        \subcaption{Mass Error ($\downarrow$)}
        \label{fig:sub_b}
    \end{subfigure}
    \caption{Performance comparison of DiffuNovo with state-of-the-art methods on mass-related metrics. Plausible prediction rate is the percentage of predicted peptides whose mass satisfies the experimental mass (higher is better). Mass error is the mean absolute error between the predicted peptide mass and the experimental mass (lower is better).} 
    \label{fig:mass_intro} 
\end{figure}
The identification of the complete set of proteins—the proteome—within a biological sample is a fundamental task in biomedicine. A comprehensive understanding of this task is critical for elucidating disease mechanisms \cite{aebersold2016mass}, discovering biomarkers \cite{geyer2017revisiting}, and identifying novel therapeutic targets for drug development \cite{moll2019target}. The principal high-throughput technology for large-scale protein analysis is tandem mass spectrometry \cite{aebersold2003mass}, which is renowned for its high sensitivity and specificity in characterizing complex biological mixtures and has revolutionized the way we study proteins on a large scale. As depicted in Figure \ref{fig:dnps}, the standard workflow \cite{zhang2013protein} begins with the enzymatic digestion of proteins from a sample into a mixture of smaller, more analytically tractable molecules called peptides. These peptides are then separated and introduced into a mass spectrometer for a two-stage analysis. In the first stage, the mass-to-charge ratio of the intact peptide, precursor ion, is measured. In the second stage, this precursor ion is isolated and fragmented, and the mass-to-charge ratios of the resulting fragment ions are measured to generate a mass spectrum, which serves as a fingerprint of the original peptide. \textbf{The central computational challenge in this workflow is solving an inverse problem: determining the amino acid sequence of a peptide from its precursor information and mass spectrum.} Successfully deciphering this mass spectrum is the crucial step that enables protein identification from the sample. To this end, de novo peptide sequencing (DNPS) offers a database-independent paradigm. It directly interprets the mass spectrum to deduce the peptide sequence from first principles, circumventing the need for a existing database. Conceptually, this task is analogous to sequence-to-sequence tasks \cite{sutskever2014sequence} in artificial intelligence, such as machine translation. In this analogy, the mass spectrum acts as the input (like a source language), which the model must translate into the target output of a peptide sequence (like a target language). 


Deep learning has been widely applied in computational biology, and a series of deep learning-based DNPS models \cite{tran2017deepnovo, qiao2021point, yilmaz2022casanovo, mao2023graphnovo, xia2024adanovo} have demonstrated significant progress. Despite these advances, existing models fail to sufficiently leverage a key principle: the mass consistency constraint. This constraint dictates that the theoretical mass of a predicted peptide sequence must match the experimentally measured precursor mass within a small tolerance. Previous methods typically frame DNPS as a multi-label classification problem, training an autoregressive model with an amino-acid-level loss function (e.g., cross-entropy). Within this framework, the precursor mass is often handled sub-optimally: it is either treated as just another numerical input feature or used merely as a post-processing filter to discard invalid candidates. This inadequate enforcement of the mass constraint leads to numerous implausible predictions. Consequently, existing DNPS models often fall short in predicting peptides that are plausible with respect to their experimental measured mass. Therefore, developing a method that leverages powerful generative models like diffusion to enforce this mass constraint is crucial for advancing DNPS.

To address this limitation, we introduce DiffuNovo, a novel regressor-guided diffusion model for de novo peptide sequencing that provides explicit, peptide-level mass control. DiffuNovo is built on a non-autoregressive Transformer architecture and comprises three main modules: a Spectrum Encoder, a Peptide Decoder, and a Peptide Mass Regressor. The Spectrum Encoder encodes the input mass spectrum into embeddings. Conditioned on this spectral embedding, the Peptide Decoder operates during the reverse diffusion process to iteratively denoise a Gaussian noise vector, passing through a series of intermediate latent variables to ultimately produce a clean latent representation of the predicted peptide sequence. The core innovation lies in the Peptide Mass Regressor, which guides the intermediate latent variables throughout this reverse process to enforce mass consistency. Specifically, we integrate the mass constraint at two critical stages: during training, a novel peptide-level mass objective is introduced to train the Regressor to predict the mass corresponding to intermediate latent variables; during inference, the pre-trained Regressor provides guidance to the Peptide Decoder by steering the generation process with gradient-based updates applied to the intermediate latent space. This compels that the final predicted peptide's mass adheres to the mass constraint. Comprehensive evaluations on established benchmarks demonstrate that DiffuNovo surpasses state-of-the-art methods in DNPS accuracy. More importantly, as shown in Figure \ref{fig:mass_intro}, DiffuNovo achieves a significant reduction in mass error compared to baseline models and thereby producing more physically plausible peptides. These innovations represent a substantial advancement toward reliable and broadly applicable DNPS.

\noindent In summary, our core contributions are as follows:
\begin{itemize}
    \item As the first DNPS model to feature explicit mass control, DiffuNovo effectively imposes this critical mass constraint throughout both the training and inference stages.
    \item We propose DiffuNovo, a novel regressor-guided diffusion model for de novo peptide sequencing that provides explicit mass control. To our knowledge, it is the first DNPS model to utilize diffusion as its core architecture.
    \item Comprehensive evaluations demonstrate that DiffuNovo achieves state-of-the-art DNPS accuracy and, through Regressor guidance, significantly reduces mass error to predict more physically plausible peptides (Figure \ref{fig:mass_intro}).
\end{itemize}




\section{Related Works}
\textbf{De Novo Peptide Sequencing (DNPS)} With the advent and prosperity of deep learning, a new wave of DNPS methods has emerged, achieving significant performance gains. DeepNovo \cite{tran2017deepnovo} was a pioneering work that first applied deep neural networks to DNPS. Subsequent research has leveraged a variety of advanced architectures, including Geometric Deep Learning \cite{qiao2021point, mao2023graphnovo}, and Transformer architecture \cite{yilmaz2022casanovo,xia2024adanovo}. Notably, while InstaNovo+ \cite{eloff2023instanovo} utilized a diffusion model for the refinement of predicted peptides, it was employed only as a post-processing step instead of the core backbone. To our knowledge, DiffuNovo is the first DNPS model to use diffusion as its core architecture.


\noindent\textbf{Diffusion Models for Controllable Generation}  Diffusion models have emerged as a prominent class of generative models, renowned for their ability to synthesize high-fidelity and fine-grained controllable samples. Recent studies have demonstrated their remarkable performance, not only in continuous domains \cite{rombach2022high,ho2022imagen,liu2023audioldm}, but also in discrete domains \cite{li2022diffusion,xu2022geodiff,watson2023novo}. Several distinct strategies have been established to implement controllability, including: incorporating feedback from an external function \cite{dhariwal2021diffusion}, training the diffusion model to accept conditioning prompt \cite{rombach2022high}, and directly modifying the denoising predictions \cite{zhang2023adding}. In contrast to the aforementioned works, our research pioneers the application to the DNPS.

\section{Preliminary}

This paper addresses the problem of de novo peptide sequencing (DNPS), which aims to determine the amino acid sequence of a peptide given its experimental mass spectrum and precursor information. The \textbf{mass spectrum} is a set of peaks $\mathbf{s} = \{{s_i}\}_{i=1}^{M} = \{ (m_i, I_i) \}_{i=1}^{M}$, where each peak $s_i = (m_i, I_i)$ consists of a mass-to-charge ratio $m_i \in \mathbb{R}$ and its corresponding intensity $I_i \in \mathbb{R}$. The \textbf{precursor} information is a tuple $\mathbf{p} = (m_{\text{prec}}, c_{\text{prec}})$, where $m_{\text{prec}} \in \mathbb{R}$ is the mass-to-charge of the precursor and $c_{\text{prec}} \in Z^{+}$ is its charge. The target output is a \textbf{peptide} $\mathbf{y}$, which is a sequence of amino acids $\mathbf{y} = (\mathbf{y}_1, \mathbf{y}_2, \dots, \mathbf{y}_N)$. Each amino acid $\mathbf{y}_i$ belongs to a pre-defined vocabulary of amino acids, $\mathbb{AA}$. The number of peaks $M$ and the peptide length $N$ is variable. The \textbf{experimentally measured mass} of the peptide $m_{exp}$ is calculated from the precursor $\mathbf{p}$ following \cite{aebersold2003mass}. The \textbf{theoretical mass} $m_{pred}$ of predicted peptide is the sum of the masses of its constituent amino acids. 

Formally, the goal of deep learning-based DNPS is learning model with parameters $\theta$ that estimates $p(\mathbf{y} | \mathbf{s}, m_{exp};\theta)$.

\section{Method}

\begin{figure*}[t]
\centering
\includegraphics[width=0.7\textwidth]{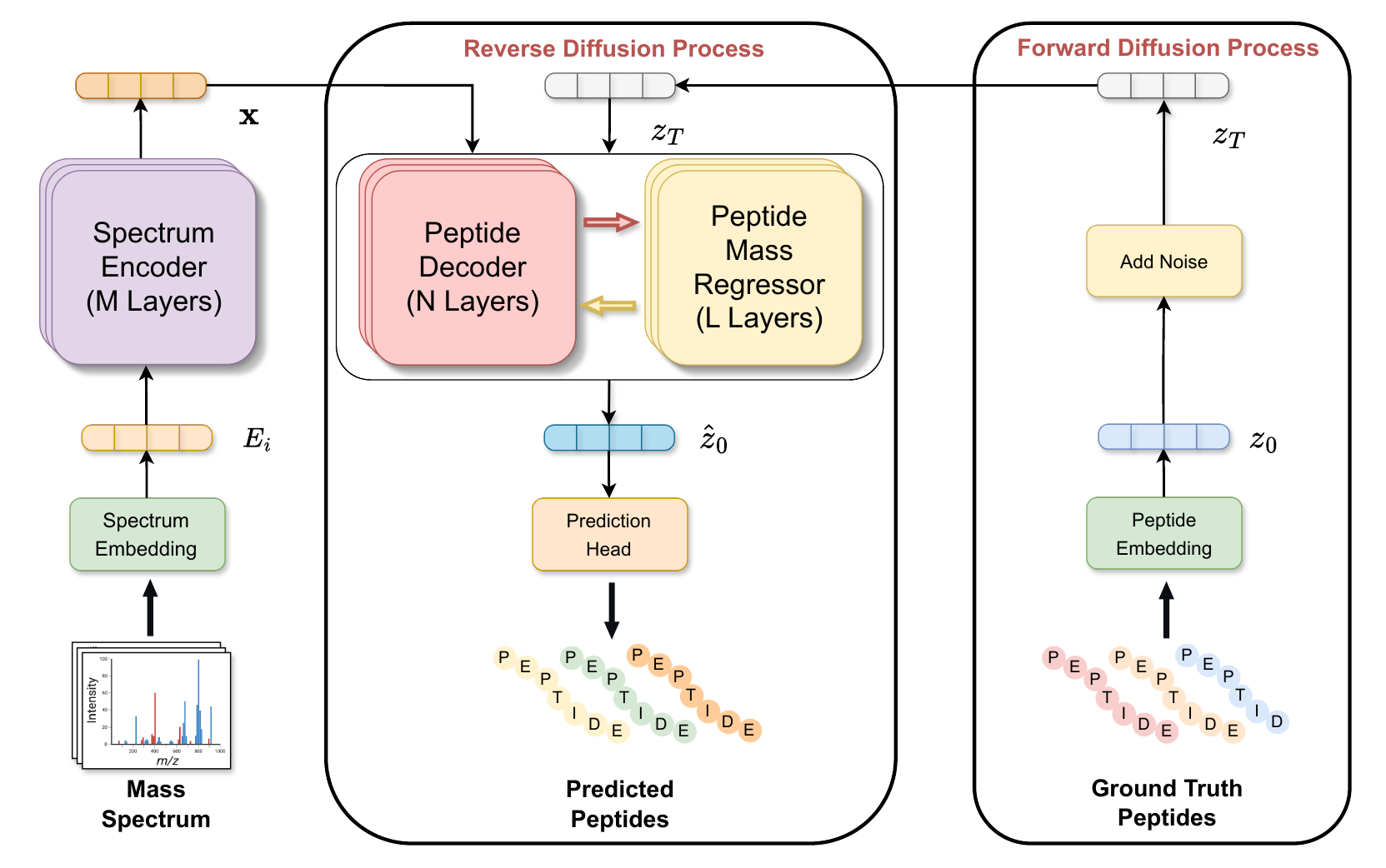}
\caption{The architecture of DiffuNovo. In the Forward Diffusion Process, a ground-truth peptide sequence is converted into an embedding $z_0$ and then corrupted by adding noise over $T$ timesteps to produce a noisy latent variable $z_T$. DiffuNovo is trained to reverse this process. The Spectrum Encoder transforms the mass spectrum into embedding $\mathbf{x}$. The core of DiffuNovo is Reverse Diffusion Process where the Peptide Decoder denoise the noisy latent $z_t$ into a cleaner latent $z_{t-1}$ conditioned on $\mathbf{x}$. Critically, the Peptide Mass Regressor provides guidance to the Peptide Decoder at each timestep $t$, compelling the prediction adheres to the mass constraint. Finally, the clean latent $\hat{z}_0$ is passed to a prediction head to output the final predicted peptide.}
\label{fig:framework}
\end{figure*}

We consider the setting of de novo peptide sequencing (DNPS) with explicit mass control. To render this complex problem more tractable without loss of generality, we decompose it into two simpler sub-problems: 1) we train a base DNPS model $p(\mathbf{y} | \mathbf{s}; \theta_1)$ on labeled dataset. 2) for explicit mass control, we train a Regressor, $p(m_{exp} | \mathbf{y}; \theta_2)$, on the same dataset to predict the mass $m_{\text{exp}}$ of a peptide sequence $\mathbf{y}$ given its high-dimensional latent variable. The goal of DiffuNovo is to utilize these two blocks to approximately sample from $p(\mathbf{y} | \mathbf{s}, m_{{exp}}, \theta)$ via Bayes rule:
\begin{equation}
\label{eq:simple_control}
p(\mathbf{y} \mid \mathbf{s}, m_{{exp}}, \theta) \propto p(\mathbf{y} \mid \mathbf{s}; \theta_1) \cdot p(m_{{exp}} \mid \mathbf{y}; \theta_2).
\end{equation}
Intuitively, the first term $p(\mathbf{y} \mid \mathbf{s}; \theta_1)$ encourages the predicted peptides $\mathbf{y}$ to be consistent with the mass spectrum $\mathbf{s}$. The second term $p(m_{{exp}} \mid \mathbf{y}; \theta_2)$ acts as a guidance to compel the predicted sequence $\mathbf{y}$ fulfills the mass $m_{exp}$.

The framework of our proposed model, DiffuNovo, is illustrated in Figure \ref{fig:framework}. It comprises three core components based on the Transformer architecture: a \textbf{Spectrum Encoder}, a \textbf{Peptide Decoder}, and a \textbf{Peptide Mass Regressor}. First, the Spectrum Encoder generates an embedding of the input mass spectrum $\mathbf{s}$. Conditioned on this embedding, the Peptide Decoder models $p(\mathbf{y} \mid \mathbf{s}; \theta_1)$ by progressively denoising a random Gaussian noise into a latent variable of the peptide sequence during the reverse diffusion process. Working in tandem with the Peptide Decoder, the Peptide Mass Regressor models $p(m_{{exp}} \mid \mathbf{y}; \theta_2)$. Regressor assesses the mass consistency of intermediate latent variables and provides guidance to steer the generation, compelling the final peptide adheres to the experimental measured mass.

\subsection{Encoding of Input by Spectrum Encoder}

Initially, DiffuNovo transforms mass spectrum $\mathbf{s}$ into a sequence of high-dimensional vectors $\{E_i\}_{i=1}^M$, suitable for processing by Transformer-based Spectrum Encoder. We follow widely used methods \citep{yilmaz2022casanovo}, involves independent vectorization of each peak in mass spectrum. 

Each peak $s_i = (m_i, I_i)$ in the mass spectrum $\mathbf{s} = \{{s_i}\}_{i=1}^{M}$ is individually mapped to a $d$-dimensional embedding, $E_i$. This transformation is achieved through two parallel pathways that separately encode the mass-to-charge ratio $m_i$ and the intensity $I_i$. The $m_i$ is encoded using a sinusoidal positional function, analogous to its use in natural language processing \cite{vaswani2017attention}, to capture the precise location of peaks within the mass domain. The intensity is projected into the embedding space via a trainable linear layer $\mathbf{W}$. The final peak embedding 
$E_i$ is the element-wise sum of these two vectors. The formal definitions are as follows:
\begin{equation}
\small
\label{eq:your_equation_label}
\begin{split}
E_i^{mz} = \Biggl[ & \sin\frac{m_i}{N_1 N_2^\frac{2}{d}}, \sin\frac{m_i}{N_1 N_2^\frac{4}{d}}, \dots, \sin\frac{m_i}{N_1 N_2^\frac{d}{d}}, \\
                 & \cos\frac{m_i}{N_1 N_2^\frac{d+2}{d}}, \cos\frac{m_i}{N_1 N_2^\frac{d+4}{d}}, \dots, \cos\frac{m_i}{N_1 N_2^2} \Biggr]
\end{split}
\end{equation}

\begin{equation}
E_i^{I} = \mathbf{W} I_i
\end{equation}

\begin{equation}
E_i = E_i^{I} + E_i^{mz}
\end{equation}

where \( d \) is the embedding dimension, and $\mathbf{W} \in \mathbb{R}^{d \times 1}$ is a trainable linear layer, \( N_1 \) and \( N_2 \) are pre-defined scalars.

The resulting sequence of peak embeddings, $\{E_i\}_{i=1}^M$, serves as the input to the Spectrum Encoder. This module employs a multi-head attention mechanism \cite{vaswani2017attention} to compute mass spectrum embedding $\mathbf{x}$, which encapsulates a holistic representation of the mass spectrum $\mathbf{s}$. Embedding $\mathbf{x}$ is then input to the Peptide Decoder.

\subsection{Training Stage of DiffuNovo}
\label{sec:train}
The training process of our proposed DiffuNovo operates in two process. The \textbf{Forward Diffusion Process} is a fixed procedure used during training where, at a given timestep, noise is added to the ground-truth peptide embeddings to yield a noisy intermediate latent variable. The \textbf{Reverse Diffusion Process} predicts peptide sequences from this noisy latent variable, conditioned on a mass spectrum embedding.

\subsubsection{Forward Diffusion Process} 
For the training of DiffuNovo, we define a specific forward process that constructs a trajectory of latent variables, $\{z_t\}_{t=0}^T$, where $T$ is maximum diffusion timestep. The initial transition from the discrete amino acid tokens of $\mathbf{y}$ to a continuous latent variable $z_0$ is defined by a conditional distribution: 
\begin{equation} \label{eq:initial_transition} q_{\phi}(z_0 \mid \mathbf{y}) = \mathcal{N}(z_0; \mathrm{Emb}_{\phi}(\mathbf{y}), (1-\alpha_0)\mathbf{I}) 
\end{equation} 

Here, $\mathrm{Emb}_{\phi}(\mathbf{y})$ is a learnable function with parameter $\phi$ that maps the discrete peptide sequence $\mathbf{y}$ into a continuous vector, and $\alpha_0$ is a predefined variance schedule parameter.

Subsequently, the variable $z_0$ is gradually perturbed until it converges to a standard Gaussian noise. The process at each intermediate timestep $t\in[1, T]$ can be formalized as:
\begin{equation}
q(z_t \mid z_0) = \mathcal{N}(z_{t};\sqrt{\bar{\alpha}_t} z_0, (1-\bar{\alpha}_t) \textbf{I}),
\label{equ:forwarddiff}
\end{equation}
where, $\bar{\alpha}_{t}=\prod_{i=1}^t\alpha_i$, and $\alpha_i$ is a noise coefficient that decreases with timestep $t$, $z_t$ is intermediate latent variable.

\subsubsection{Reverse Diffusion Process} 

The reverse denoising process is designed to invert the forward diffusion by learning a parameterized transition, $p_\theta(z_{t-1} \mid z_t, t, \mathbf{x})$, that progressively removes noise conditioned on the mass spectrum embedding $\textbf{x}$. This process starts with a sample $z_T$ from a standard Gaussian distribution, $z_T \sim \mathcal{N}(\mathbf{0}, \mathbf{I})$, and iteratively applies the denoising step until $t=0$. Each transition in this reverse Markov chain is modeled as a Gaussian distribution: 
\begin{equation} 
\label{equ:reversediff} 
p_\theta(z_{t-1}\mid z_t, t, \mathbf{x}) = \mathcal{N}\big(z_{t-1}; {\mu_{\theta}}(z_t, t, \mathbf{x}), {\Sigma_{\theta}}(z_t, t, \mathbf{x})\big), 
\end{equation}

The variance $\Sigma_{\theta}$ is kept fixed to a predefined schedule following \citep{li2022diffusion}.
To parameterizing the mean $\mu_{\theta}$ , we train the Peptide Decoder denoted as $\hat{z}_0 = \mathbf{g}_\theta(z_t, t, \mathbf{x})$, to predict the clean embedding $\hat{z}_0$ from the noisy latent variable $z_t$. This non-autoregressive prediction of the final state is a common and effective strategy in diffusion models \citep{lin2023text}. The required mean $\mu_{\theta}$ for Equation \ref{equ:reversediff} can then be calculated in a closed form based on the predicted $\hat{z}_0$ and the current state $z_t$ following \cite{ho2020denoising}.

The final step of the reverse diffusion process maps the fully denoised latent variable $\hat{z_0}$ (continuous embedding) to the target peptide sequence $\mathbf{{y}}$ (discrete sequence). This is achieved through a prediction head that models the probability of each amino acid in each position $i$ independently: 
\begin{equation} 
\label{equ:rounding_step} 
p_{\theta}(\mathbf{y} \mid \hat{z_0}) = \prod_{i=1}^{N} p_{\theta}(\mathbf{y}_i \mid \hat{z}_0^i) 
\end{equation}
The DiffuNovo's parameters are optimized by maximizing the evidence lower bound (ELBO) of the log-likelihood \cite{ho2020denoising}. This yields a simplified and effective training objective, which is a combination of a reconstruction term and a denoising-matching term: 
\begin{equation} 
\small
\label{equ:difflmloss}
\begin{split}
 \mathcal{L}_{\text{1}} &= \mathcal{L}_{\text{Peptide Decoder}} \\
    &= \mathbb{E}_{q_{\phi}(z_{0:T} \mid \mathbf{y})} \bigg[ \sum_{t=1}^T  \Vert z_0-\mathbf{g}_\theta(z_t,t,\mathbf{x}) \Vert^2-\log p_{\theta}(\mathbf{y} \mid z_0) \bigg]
\end{split}
\end{equation}

In parallel with the denoising task, we introduce and co-train the Peptide Mass Regressor. The function of this Regressor is to predict the peptide mass $m_{pred}$ directly from the noisy intermediate latent variable $z_t$ at any timestep $t$:

\begin{equation}
\label{equ:mass_regressor}
{m_{pred}}(z_t, \theta) = \sum_{i=1}^{N} \sum_{\mathbf{y}_j \in \mathbb{AA}} p_{\theta}(\mathbf{y}_j \mid z_t^i, \theta) \cdot m(\mathbf{y}_j)
\end{equation}
where $\mathbb{AA}$ is the set of amino acids and $m(\mathbf{y}_i)$ is the mass of amino acid $\mathbf{y}_i$. Regressor is trained on the same dataset as the Peptide Decoder, and its parameters are optimized using mean squared error (MSE) loss between the predicted mass $m_{pred}$ and the experimental measured mass $m_{exp}$:

\begin{equation}
\label{equ:mass_loss}
\mathcal{L}_{\text{2}} =\mathcal{L}_{\text{Regressor}} = \mathbb{E}_{q_{\phi}(z_{0:T} \mid \mathbf{y})} \| {m_{pred}}(z_t, \theta) - m_{exp})\|^2
\end{equation}

\begin{algorithm}[t]
\caption{Training Stage of DiffuNovo.}
\label{alg:train}
\hspace*{\algorithmicindent} 
\textbf{Input}: Labeled Dataset $\mathcal{D}=\{(\textbf{s}, m_{exp}), \textbf{y}\}$, maximum diffusion timestep $T$ and maximum peptide length $N$. \\
\hspace*{\algorithmicindent} 
\textbf{Output}: Optimized model parameters $\theta$.

\begin{algorithmic}[1]
\REPEAT
\STATE Sample a data instance $(\mathbf{s}, m_{\text{exp}}, \mathbf{y}) \sim \mathcal{D}$.
\STATE {Encode input $\mathbf{s}=\{s_i\}_{i=1}^{M}$ into continuous representations $\{E_i\}_{i=1}^{M}$} and compute mass spectrum embedding: $\mathbf{x} \leftarrow \text{Spectrum Encoder}(\{E_i\}_{i=1}^{M})$.
\STATE Maps the discrete peptide $\mathbf{y}$ into embedding vector: 
\begin{equation}
z_0 \sim q_{\phi}(z_0 \mid \mathbf{y}) = \mathcal{N}(z_0; \mathrm{Emb}_{\phi}(\mathbf{y}), (1-\alpha_0)\mathbf{I})
\end{equation}
\STATE Sample a timestep $t \sim [1, T]$ and construct the noisy latent variable $z_{t}$ with Gaussian reparameterization:
\begin{equation}
z_{t} \sim q(z_t \mid z_0)=\mathcal{N}(z_{t}; \sqrt{\bar{\alpha}_{t}}z_0, (1-\bar{\alpha}_{t})\mathbf{I}). 
\end{equation}
\STATE According to Equation \ref{equ:difflmloss} and Equation \ref{equ:mass_loss}, employ gradient descent to optimize the objective: 
\begin{equation}
\small
\label{equ:gradient}
\begin{split}
\min_{\theta} \Biggl\{ & \Bigl[ \big\| z_0 - \mathbf{g}_{\theta}(z_{t}, t;\mathbf{x})\big\|^2 -\log p_\theta(\mathbf{y}\mid z_{0})\Bigr] \\ 
& + \Bigl\| \sum_{i=1}^{N} \sum_{\mathbf{y}_j \in \mathbb{AA}} p_{\theta}(\mathbf{y}_j \mid z_{t}^i, \theta) \cdot m(\mathbf{y}_j) - m_{\exp} \Bigr\|^2 \Biggr\}
\end{split}
\end{equation}

\UNTIL{converged}
\end{algorithmic}
\label{algorithm:1}
\end{algorithm}

\noindent The whole process of the DiffuNovo model training stage can be summarized by the pseudocode in Algorithm \ref{alg:train}.

\subsection{Inference Stage of DiffuNovo}

\begin{algorithm}[t]
\small
\caption{Inference Process of DiffuNovo.}
\label{alg:infer}
\hspace*{\algorithmicindent} 
\textbf{Input}: Inference instance $(\textbf{s}, m_{exp})$ from test dataset $\mathcal{D}=\{(\textbf{s}, m_{exp})\}$, maximum diffusion decoding timestep $T$, trained model parameters $\theta$ and gradient-based guidance step $s$, scalar coefficient $\lambda_1$ and $\lambda_2$. \\
\hspace*{\algorithmicindent} 
\textbf{Output}: Predicted peptide sequence $\mathbf{\hat{y}}$.

\begin{algorithmic}[1]
\STATE {Encode inputs $\mathbf{s}=\{s_i\}_{i=1}^{M}$ into continuous representations $\{E_i\}_{i=1}^{M}$} and compute mass spectrum embedding: $\mathbf{x} \leftarrow \text{Spectrum Encoder}(\{E_i\}_{i=1}^{M})$.
\STATE Uniformly select a decreasing subsequence of timesteps $t_{M:0}$ ranging from $T$ to $0$.
\STATE Sample $z_{t_{M}} \sim \mathcal{N}(\textbf{0}, \textbf{I})$.
\FOR{$i = M$ to $1$}

\STATE Get the current timesteps $t_i$ and
the subsequent timestep $t_{i-1}$ from the pre-defined timesteps $t_{M:0}$

\STATE Compute denoising mean through Peptide Decoder: 
\begin{equation}
{\mu}_\theta (z_{t_i}, t_i, \mathbf{x})\leftarrow \lambda_1 z_{t_i} + \lambda_2 \mathbf{g}_\theta(z_{t_i}, t_i, \mathbf{x})
\end{equation}

\STATE Compute gradient-based update for latent variables $z_{t_i}$ through Peptide Mass Regressor:
\begin{equation}
\Delta z_{t_{i}} = \nabla_{z_{t_i}} \big\|{m}_{pred}(z_{t_i}, \theta) - m_{exp}\big\|^2
\end{equation}

\STATE The subsequent latent variables $z_{t_{i-1}}$ is then sampled from Gaussian distribution: 
\begin{equation}
\begin{gathered}
p_{\theta}\big(z_{t_{i-1}}\mid z_{t_i}, t_i, \mathbf{x}, m_{exp}\big) \\ \sim \mathcal{N}\big(z_{t_{i-1}};{\mu}_\theta (z_{t_i}, t_i, \mathbf{x}) + s \Delta z_{t_{i}}, \sigma \textbf{I}\big) 
\end{gathered}
\label{equ:reverse_sample2_2}
\end{equation}
\ENDFOR
\STATE Map $z_{0}$ to the peptide sequence $\mathbf{\hat{y}}$ through prediction head.
\end{algorithmic}
\label{algorithm:2}
\end{algorithm}

\subsubsection{Reverse Process}
The inference process of DiffuNovo is summarized in Algorithm \ref{alg:infer}. During inference, DiffuNovo executes the reverse diffusion process to predict peptide sequence $\mathbf{y}$ conditioned on input mass spectrum $\mathbf{x}$. This process begins with an initial Gaussian noise vector $z_T$, which is iteratively refined through $T$ timesteps. At each step $t$, the Peptide Decoder takes the noisy latent variable $z_t$ and the spectrum embedding $\mathbf{x}$ from the Spectrum Encoder as input to predict a cleaner latent variable $z_{t-1}$:
\begin{equation}
\small
\begin{gathered}
p_{\theta}\big(z_{t-1}\mid z_{t},t, \mathbf{x}, m_{\text{exp}}\big) \sim \mathcal{N}\big(z_{t-1};{\mu}_\theta (z_t, \mathbf{x}, t) + s \Delta z_{t_{i}}, \sigma \textbf{I}\big) 
\end{gathered}
\label{equ:reverse_sample2_2}
\end{equation}

Crucially, the inference process is different from the reverse diffusion mentioned in Section~\ref{sec:train} by: 1) the denoising process begins not by adding noise to a ground-truth peptide $\mathbf{y}$, but by sampling an initial latent variable $z_T$ from a standard Gaussian distribution. 2) incorporating an explicit mass control by intermediate latent variable by the term $s\Delta z_{t_{i}}$ ($s$ is a scalar) using the trained Peptide Mass Regressor.

Our approach to explicit mass control is inspired by Equation \ref{eq:simple_control}, but instead of directly controlling the discrete peptide sequence, we control the sequence of continuous intermediate latent variables $z_{0:T}$ during reverse diffusion process. As a further refinement of the simplified formulation in Equation ~\ref{eq:simple_control}, controlling $z_{0:T}$ is equivalent to decoding from the posterior $p(z_{0:T} \mid \mathbf{x} , m_{exp}) = \prod_{t=1}^T p(z_{t-1} \mid z_{t}, \mathbf{x}, m_{exp})$, and we decompose this complex inference problem by: 
\begin{equation}
p(z_{t-1} \mid z_{t}, \mathbf{x}, m_{exp}) \propto p(z_{t-1} \mid z_{t}, \mathbf{x}) \cdot p( m_{exp} \mid z_{t-1}, z_{t})
\end{equation}

We further simplify $p(m_{exp} \mid z_{t-1}, z_{t})=p(m_{exp} \mid z_{t-1})$ via conditional independence assumptions from prior work on controlling diffusions \cite{song2020score}. Consequently, for the $t$-th timestep, we run gradient update on $z_{t-1}$:
\begin{align*}
\begin{split}
    & \nabla_{z_{t-1}} p(z_{t-1} \mid z_{t}, \mathbf{x},  m_{exp}, \theta)  \\=& \nabla_{z_{t-1}} p(z_{t-1} \mid z_{t}, \mathbf{x},  \theta_1) + \nabla_{z_{t-1}}   p(m_{exp} \mid z_{t-1}, \theta_2)
\end{split}
\end{align*}
where both $ p(z_{t-1} \mid z_{t}, \mathbf{x}, \theta_1)$ and $ p(m_{exp} \mid z_{t-1}, \theta_2)$ are differentiable: the first term is parametrized by Peptide Decoder, and the second term is parametrized by a neural network-based Peptide Mass Regressor. We run gradient updates $\nabla_{z_{t-1}}  p(m_{exp} \mid z_{t-1}, \theta_2)$ on the latent space to steer it towards fulfilling the mass consistency constrain.

\subsubsection{Final Prediction}

Finally, prediction head maps the fully denoised $\hat{z}_0$ to the predicted peptide sequence $\mathbf{y}$. Similarly to beam search \cite{freitag2017beam}, DiffuNovo predicts a set of candidates and select the final prediction by:
\begin{itemize}
    \item \textbf{DiffuNovo (Logits):} This variant selects the candidate with the highest peptide-level log-probability, as determined by the logits from the final projection head.
    \item \textbf{DiffuNovo (MBR):} This variant employs Minimum Bayes Risk (MBR) decoding \citep{kumar2004minimum} to select the optimal candidates from the predicted set.
\end{itemize}

\section{Experiments}

\subsection{Experimental Settings}

All experimental settings in this paper adhere to the NovoBench benchmark \cite{zhou2024novobench}. Our evaluation leverages three representative datasets, selected for their diverse sizes, resolutions, and biological origins: the Nine-species Dataset \citep{tran2017deepnovo}, the HC-PT Dataset \citep{eloff2023instanovo}, and the Seven-species Dataset \citep{tran2017deepnovo}.

To comprehensively evaluate the performance of DiffuNovo, we compare it against a suite of advanced baselines, including DeepNovo \citep{tran2017deepnovo}, PointNovo \citep{qiao2021point}, CasaNovo \citep{yilmaz2022casanovo}, AdaNovo \citep{xia2024adanovo}, and HelixNovo \citep{yang2024helixnovo}. The performance of all models was assessed using standard metrics: (1) Peptide-level Precision serves as the primary indicator of model performance; (2) Peptide-level Area Under the Curve (AUC) assesses performance across different confidence thresholds; and (3) Amino Acid-level Precision and Recall evaluate performance at a finer granularity.

\subsection{Experimental Results}

\subsubsection{DiffuNovo Achieves State-of-the-art Performance on Most Benchmark Metrics} The empirical results, summarized in Table \ref{tab:main} and \ref{tab:ptm_comparison}, demonstrate that DiffuNovo achieves state-of-the-art performance, consistently outperforming leading baselines across most datasets and metrics.

The performance on standard DNPS benchmarks is detailed in Table \ref{tab:main}. Our model demonstrates exceptional capabilities, particularly in peptide-level precision. The DiffuNovo(Logits) variant achieves the best peptide precision across the major datasets, with a precision of 0.572 on the 9-species dataset and 0.485 on the HC-PT dataset, decisively outperforming all baseline models. Furthermore, the DiffuNovo(MBR) variant showcases the model's comprehensive power by securing top performance across other metrics, including the highest peptide-level AUC and amino acid-level precision and recall. These results highlights the DiffuNovo's core strength in accurately predicting the correct peptide sequence for de novo peptide sequencing.

The identification of amino acids with post-translational modifications (PTMs) holds important biological significance because it plays a pivotal role \cite{deribe2010post}. As detailed in Table \ref{tab:ptm_comparison}, DiffuNovo achieves the highest PTM precision, outperforming other methods by a substantial margin. By enforcing a strict adherence to the experimental mass through Regressor, DiffuNovo effectively eliminating candidate peptides that are physically implausible. This constraint is crucial for PTMs, where subtle mass shifts differentiate PTM from canonical amino acid.

  

\subsubsection{The Peptides Predicted by DiffuNovo Exhibit A Highly Significant Enhancement in Mass Consistency} 
\begin{figure}[!ht] 
    \centering 

    \begin{subfigure}[b]{0.48\columnwidth}
        \centering
        \includegraphics[width=\linewidth]{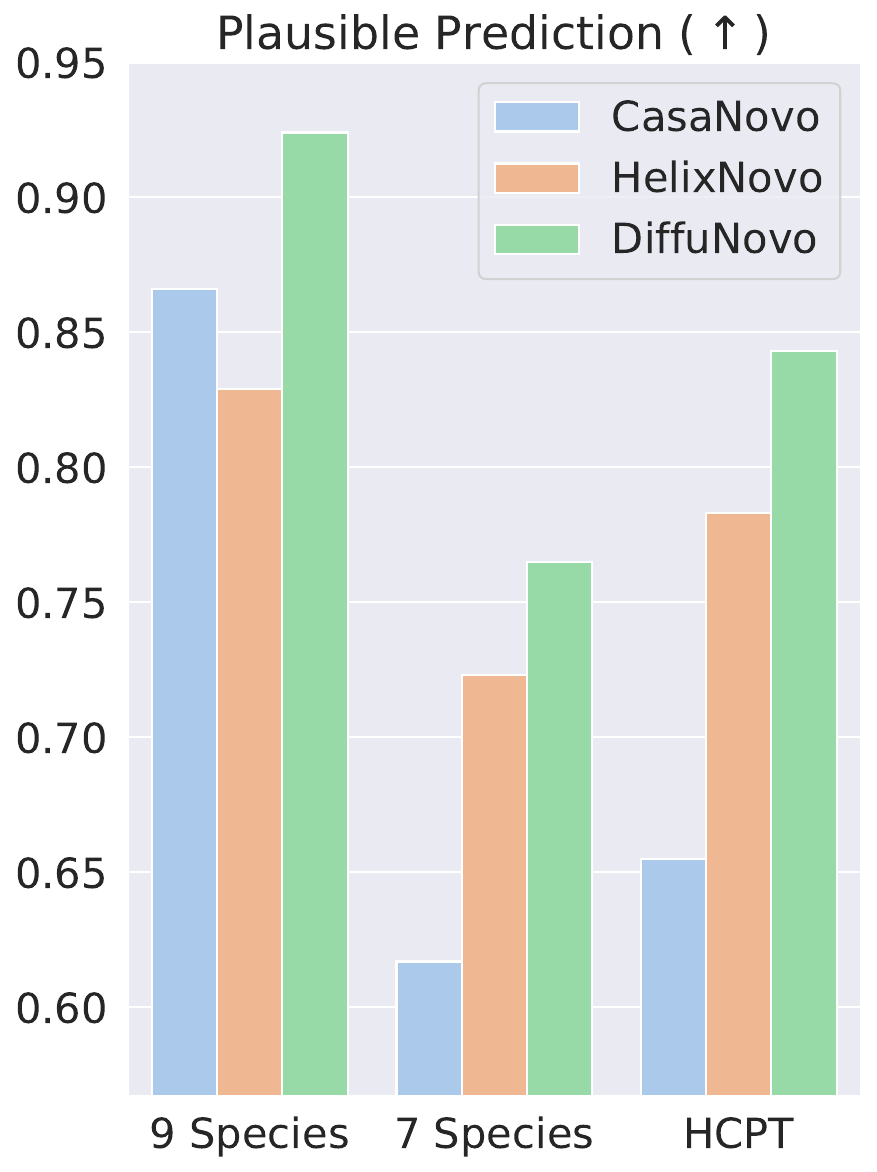} 
        \subcaption{Plausible Prediction Rate ($\uparrow$)} 
        \label{fig:sub_a} 
    \end{subfigure}
    %
    \hfill 
    %
    \begin{subfigure}[b]{0.48\columnwidth}
        \centering
        \includegraphics[width=\linewidth]{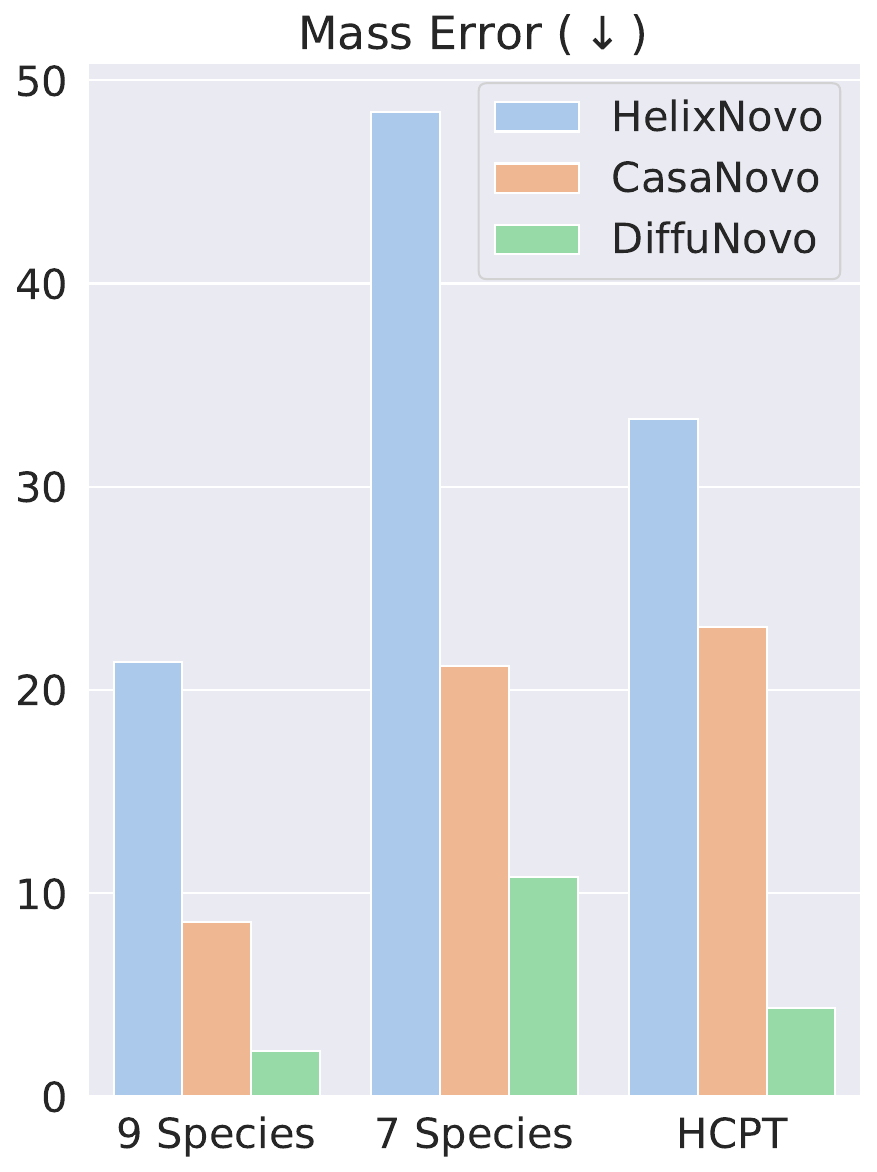}
        \subcaption{Mass Error ($\downarrow$)}
        \label{fig:sub_b}
    \end{subfigure}
    
    \caption{Performance comparison on mass-related metrics.} 
    \label{fig:mass_exp} 
\end{figure}

To quantitatively evaluate the effectiveness of our proposed explicit mass control, we compared the theoretical mass of the peptides predicted by DiffuNovo with their experimentally determined precursor mass. Table \ref{tab:mass_cons} presents the Mean Absolute Error of this mass discrepancy. Figure \ref{fig:mass_exp} illustrates the plausible prediction rate, defined as the proportion of predictions where the mass error is less than 1e-3 Da.

\begin{table}[!h]
\centering
\begin{tabular}{l p{1.5cm} p{1.5cm} p{1.5cm}}
\toprule
\textbf{Method} & {\textbf{HelixNovo}} & {\textbf{CasaNovo}} & {\textbf{DiffuNovo}} \\
\midrule
MAE($\downarrow$)     & 21.388 & 8.583 & \textbf{2.251} \\
\midrule[1.2pt]
\multicolumn{3}{c}{\textbf{Analysis of DiffuNovo's Improvement}} \\
\midrule
\textbf{DiffuNovo vs.} & \textbf{HelixNovo} & \textbf{CasaNovo} & \textbf{DiffuNovo} \\
\midrule
X\% Decrease & 89.5\% & 73.8\% & 0\% \\
Fold Decrease & 9.50$\times$ & 3.81$\times$ & 1.00$\times$ \\
\bottomrule
\end{tabular}
\caption{Evaluation of mass error. We compares DiffuNovo with two baselines and reports the Mean Absolute Error (MAE), where lower is better. The upper section shows the mass errors. The lower section quantifies this improvement.}
\label{tab:mass_cons}
\end{table}

As Table \ref{tab:mass_cons}, DiffuNovo achieves a MAE of 2.251, representing a remarkable 89.5\% (a 9.50-fold decrease) and 73.8\% (a 3.81-fold decrease) reduction compared to HelixNovo and CasaNovo. This empirical evidence strongly validates that the explicit mass control is highly effective.

As illustrated in Figure \ref{fig:mass_exp}, DiffuNovo consistently outperforms state-of-the-art baselines on key mass-related metrics. Figure \ref{fig:sub_a} shows that our model achieves a substantially higher plausible prediction rate, indicating that a greater proportion of prediction are physically plausible as their theoretical mass aligns with the experimental mass. Concurrently, Figure \ref{fig:sub_b} reveals a dramatic reduction in mass error.

\begin{table*}[htbp]
\centering
\begin{tabularx}{\textwidth}{l|cc|cc|cc|cc|cc|cc}
\toprule
 & \multicolumn{6}{c|}{Peptide-level Performance} & \multicolumn{6}{c}{Amino Acid-level Performance} \\
\cmidrule(lr){2-13}
 Models & \multicolumn{2}{c|}{9-species} & \multicolumn{2}{c|}{HC-PT} & \multicolumn{2}{c|}{7-species} & \multicolumn{2}{c|}{9-species} & \multicolumn{2}{c|}{HC-PT} & \multicolumn{2}{c}{7 species} \\
  & Prec. & AUC & Prec. & AUC & Prec. & AUC & Prec. & Recall & Prec. & Recall & Prec. & Recall \\
\midrule
DeepNovo & 0.428 & 0.376 & 0.313 & 0.255 & 0.204 & 0.136 & 0.696 & 0.638 & 0.531 & 0.534 & \textbf{0.492} & 0.433 \\
PointNovo & 0.480 & 0.436 & \underline{0.419} & \underline{0.373} & 0.022 & 0.007 & {0.740} & 0.671 & \underline{0.623} & \underline{0.622} & 0.196 & 0.169\\
CasaNovo & 0.481 & 0.439 & 0.211 & 0.177 & 0.119 & 0.084 & 0.697 & 0.696 & 0.442 & 0.453 & 0.322 & 0.327 \\
AdaNovo & 0.505 & \underline{0.469} & 0.212 & 0.178 & 0.174 & 0.135 & 0.698 & 0.709 & 0.442 & 0.451 & 0.379 & 0.385 \\
HelixNovo  & \underline{0.517} & {0.453} & {0.356} & {0.318} & \textbf{0.234} & \textbf{0.173} & \underline{0.765} & \underline{0.758} & {0.588} & {0.582} & \underline{0.481} & \textbf{0.472}\\
\midrule
{{DiffuNovo(Logits)}} & \textbf{{0.572}} & {0.413} & \textbf{{0.485}} & {0.324} & \underline{0.233} & 0.104 & {0.785} & {0.783} & {0.648} & {0.648} & 0.430 & 0.428\\
{{DiffuNovo(MBR)}} & {0.565} & \textbf{{0.536}} & {0.458} & \textbf{0.434} & 0.193 & \underline{0.162} & \textbf{{0.791}} & \textbf{{0.789}} & \textbf{{0.654}} & \textbf{0.654} & 0.437 & \underline{0.435}\\

\bottomrule
\end{tabularx}
\caption{The comparison of de novo peptide sequencing performance between our proposed model, DiffuNovo, and other state-of-the-art methods on the three benchmark datasets. We report precision and AUC at the peptide level, and precision and recall at the amino acid level. The best and the second best are highlighted with \textbf{bold} and \underline{underline}, respectively.}
\label{tab:main}
\end{table*}

\begin{table}[t] 
  \captionsetup{labelsep=period, justification=justified, singlelinecheck=false}
  \centering 
  
  \begin{tabular}{lccc} 
    \toprule
    & \multicolumn{3}{c}{PTM Precision} \\
    \cmidrule(lr){2-4}
    Models      & Nine-Species & HC-PT & Seven-Species \\
    \midrule
    DeepNovo  & 0.576 & 0.626 & 0.391 \\
    PointNovo  & 0.629 & \underline{0.676} & 0.117 \\
    AdaNovo    & 0.652 & 0.552 & 0.448\\
    CasaNovo    & \underline{0.706} & 0.501 & 0.360 \\
    HelixNovo    & 0.680 & 0.568 & \underline{0.473}\\
    \midrule
    \textbf{DiffuNovo} & \textbf{0.822} & \textbf{0.705} & \textbf{0.515}\\
    \bottomrule
  \end{tabular}
  \caption{Empirical comparison of PTM identification. We evaluate the ability of DiffuNovo and other models to identify Post-Translational Modifications (PTMs) on the benchmark datasets. The best results and the second best are highlighted with \textbf{bold} and \underline{underline}, respectively.}
  \label{tab:ptm_comparison}
\end{table}

\begin{figure}[!t]
    \centering
    \includegraphics[width=0.8\linewidth]{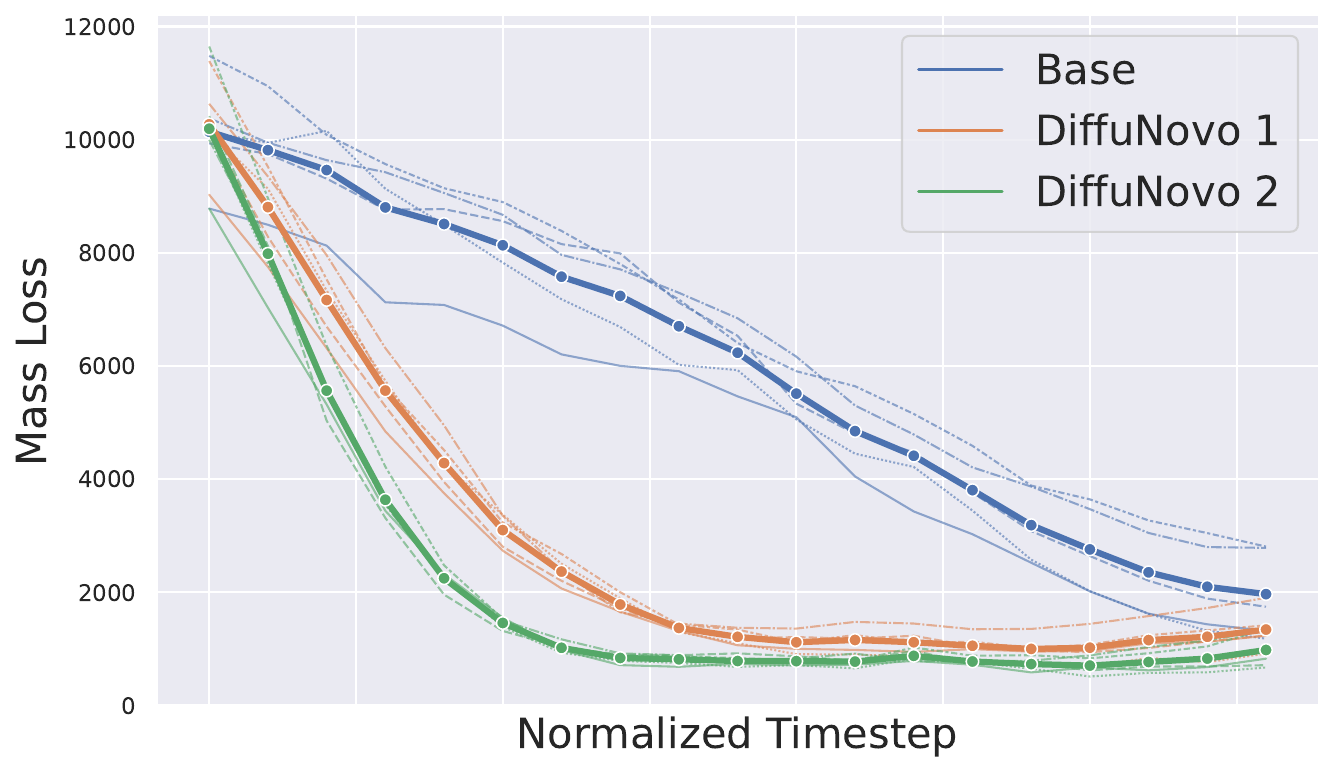}
    \caption{Mass error loss curves. The figure illustrates the mass error loss (MSE) as a function of the normalized diffusion timestep. We conducted 
    experiment on 5 random batches (2560 samples), calculating and plotting the loss for each batch (dashed line) and the average loss (solid line).}
    \label{fig:mass_curve}
\end{figure}

This enhanced performance is not coincidental but is intrinsically linked to the core design of DiffuNovo: by integrating a Regressor to guide in the latent space, our model actively steers the diffusion process towards peptides whose theoretical mass aligns with the experimental mass.

 

\subsubsection{The Significant Reduction in Mass Error is Directly Attributable to the Regressor-based Guidance} 

  

To validate the effectiveness of Regressor for reduction in mass error, we analyzed the mass error trajectory throughout the reverse diffusion process. Figure \ref{fig:mass_curve} illustrates three variants of DiffuNovo: a baseline version without Regressor (Base), and two guided versions with different guidance steps (DiffuNovo 1 with step=5e-3 and DiffuNovo 2 with step=1e-2).



The results provide compelling evidence that the regressor-based guidance is a critical component for achieving reduction in mass error. As depicted by the blue curve, the original unguided model reduces the mass error gradually, but its final error remains substantial. In stark contrast, the regressor-based variants leads to a markedly low error. Both DiffuNovo 1 (orange curve) and DiffuNovo 2 (green curve) demonstrate a significantly faster and deeper reduction in mass error from the very early stages of the reverse diffusion process. This visually confirms that the Regressor effectively steers the diffusion towards states that are consistent with the target experimental measured mass.

\subsubsection{Ablation Study}

In the ablation study, we investigated the contribution of each key component within DiffuNovo. We removed the Spectrum Encoder, the Peptide Decoder, and the Peptide Mass Regressor, then trained and evaluated these ablated models on the HC-PT dataset. All other settings were kept identical. The results is summarized in Table \ref{tab:ablation}, lead to the conclusion that each module is crucial.

\begin{table}[htbp]
\centering
\small
\begin{tabular}{l|cccc}
\toprule
{\textbf{Model}} & \multicolumn{2}{c}{\textbf{Performance}} & \multicolumn{2}{c}{\textbf{Mass Constraint}} \\
\cmidrule(lr){2-3} \cmidrule(lr){4-5}
 & Peptide & AA & Plausible & MAE \\
  & Prec.($\uparrow$) & Prec.($\uparrow$) & Rate($\uparrow$) & Value($\downarrow$) \\
\midrule
Full Model          & \textbf{0.485} & \textbf{0.648} & \textbf{0.843} & \textbf{4.350} \\
- w/o Encoder & 0.104 & 0.313 & 0.686 & 16.509 \\
- w/o Decoder  & 0.066 & 0.267 & 0.652 & 35.260 \\
- w/o Regressor & 0.437 & 0.641 & 0.703 & 14.555 \\
\bottomrule
\end{tabular}
\caption{Ablation study of the model components.} 
\label{tab:ablation}
\end{table}

\section{Conclusion}
In this paper, we identify that previous DNPS methods often handle mass information in trivial manner, leading to implausible predictions that are inconsistent with the experimental mass. To address this limitation, we introduce DiffuNovo, a novel regressor-guided diffusion model that provides explicit mass control. Guidance from Regressor via gradient-based updates in the latent space compels the predicted peptides adhere to the mass constraint. Comprehensive evaluations demonstrate that DiffuNovo surpasses state-of-the-art methods in DNPS accuracy. We show that the Regressor's guidance significantly reduces the mass error and increases the rate of plausible predictions. These innovations demonstrate that DiffuNovo model has achieved a substantial advancement toward more robust and reliable DNPS.

\section{Acknowledgments}
We sincerely thank the anonymous reviewers for their insightful comments and constructive suggestions. This research is supported by National Natural Science Foundation of China Project (No. 623B2086), TeleAI of China Telecom, Tencent and Ant Group

\bibliography{main}

\begin{thebibliography}{29}
\providecommand{\natexlab}[1]{#1}

\bibitem[{Aebersold and Mann(2003)}]{aebersold2003mass}
Aebersold, R.; and Mann, M. 2003.
\newblock Mass spectrometry-based proteomics.
\newblock \emph{Nature}, 422(6928): 198--207.

\bibitem[{Aebersold and Mann(2016)}]{aebersold2016mass}
Aebersold, R.; and Mann, M. 2016.
\newblock Mass-spectrometric exploration of proteome structure and function.
\newblock \emph{Nature}, 537(7620): 347--355.

\bibitem[{Deribe, Pawson, and Dikic(2010)}]{deribe2010post}
Deribe, Y.~L.; Pawson, T.; and Dikic, I. 2010.
\newblock Post-translational modifications in signal integration.
\newblock \emph{Nature structural \& molecular biology}, 17(6): 666--672.

\bibitem[{Dhariwal and Nichol(2021)}]{dhariwal2021diffusion}
Dhariwal, P.; and Nichol, A. 2021.
\newblock Diffusion models beat gans on image synthesis.
\newblock \emph{Advances in neural information processing systems}, 34: 8780--8794.

\bibitem[{Eloff et~al.(2023)Eloff, Kalogeropoulos, Morell, Mabona, Jespersen, Williams, van Beljouw, Skwark, Laustsen, Brouns et~al.}]{eloff2023instanovo}
Eloff, K.; Kalogeropoulos, K.; Morell, O.; Mabona, A.; Jespersen, J.~B.; Williams, W.; van Beljouw, S.~P.; Skwark, M.; Laustsen, A.~H.; Brouns, S.~J.; et~al. 2023.
\newblock De novo peptide sequencing with InstaNovo: Accurate, database-free peptide identification for large scale proteomics experiments.
\newblock \emph{bioRxiv}, 2023--08.

\bibitem[{Freitag and Al-Onaizan(2017)}]{freitag2017beam}
Freitag, M.; and Al-Onaizan, Y. 2017.
\newblock Beam search strategies for neural machine translation.
\newblock \emph{arXiv preprint arXiv:1702.01806}.

\bibitem[{Geyer et~al.(2017)Geyer, Holdt, Teupser, and Mann}]{geyer2017revisiting}
Geyer, P.~E.; Holdt, L.~M.; Teupser, D.; and Mann, M. 2017.
\newblock Revisiting biomarker discovery by plasma proteomics.
\newblock \emph{Molecular systems biology}, 13(9): 942.

\bibitem[{Ho et~al.(2022)Ho, Chan, Saharia, Whang, Gao, Gritsenko, Kingma, Poole, Norouzi, Fleet et~al.}]{ho2022imagen}
Ho, J.; Chan, W.; Saharia, C.; Whang, J.; Gao, R.; Gritsenko, A.; Kingma, D.~P.; Poole, B.; Norouzi, M.; Fleet, D.~J.; et~al. 2022.
\newblock Imagen video: High definition video generation with diffusion models.
\newblock \emph{arXiv preprint arXiv:2210.02303}.

\bibitem[{Ho, Jain, and Abbeel(2020)}]{ho2020denoising}
Ho, J.; Jain, A.; and Abbeel, P. 2020.
\newblock Denoising diffusion probabilistic models.
\newblock \emph{Advances in neural information processing systems}, 33: 6840--6851.

\bibitem[{Kumar and Byrne(2004)}]{kumar2004minimum}
Kumar, S.; and Byrne, W. 2004.
\newblock Minimum bayes-risk decoding for statistical machine translation.
\newblock Technical report, JOHNS HOPKINS UNIV BALTIMORE MD CENTER FOR LANGUAGE AND SPEECH PROCESSING (CLSP).

\bibitem[{Li et~al.(2022)Li, Thickstun, Gulrajani, Liang, and Hashimoto}]{li2022diffusion}
Li, X.; Thickstun, J.; Gulrajani, I.; Liang, P.~S.; and Hashimoto, T.~B. 2022.
\newblock Diffusion-lm improves controllable text generation.
\newblock \emph{Advances in neural information processing systems}, 35: 4328--4343.

\bibitem[{Lin et~al.(2023)Lin, Gong, Shen, Wu, Fan, Lin, Duan, and Chen}]{lin2023text}
Lin, Z.; Gong, Y.; Shen, Y.; Wu, T.; Fan, Z.; Lin, C.; Duan, N.; and Chen, W. 2023.
\newblock Text Generation with Diffusion Language Models: A Pre-training Approach with Continuous Paragraph Denoise.
\newblock arXiv:2212.11685.

\bibitem[{Liu et~al.(2023)Liu, Chen, Yuan, Mei, Liu, Mandic, Wang, and Plumbley}]{liu2023audioldm}
Liu, H.; Chen, Z.; Yuan, Y.; Mei, X.; Liu, X.; Mandic, D.; Wang, W.; and Plumbley, M.~D. 2023.
\newblock Audioldm: Text-to-audio generation with latent diffusion models.
\newblock \emph{arXiv preprint arXiv:2301.12503}.

\bibitem[{Mao et~al.(2023)Mao, Zhang, Xin, and Li}]{mao2023graphnovo}
Mao, Z.; Zhang, R.; Xin, L.; and Li, M. 2023.
\newblock Mitigating the missing-fragmentation problem in de novo peptide sequencing with a two-stage graph-based deep learning model.
\newblock \emph{Nature Machine Intelligence}, 5(11): 1250--1260.

\bibitem[{Moll and Colombo(2019)}]{moll2019target}
Moll, J.; and Colombo, R. 2019.
\newblock \emph{Target identification and validation in drug discovery}.
\newblock Springer.

\bibitem[{Qiao et~al.(2021)Qiao, Tran, Xin, Chen, Li, Shan, and Ghodsi}]{qiao2021point}
Qiao, R.; Tran, N.~H.; Xin, L.; Chen, X.; Li, M.; Shan, B.; and Ghodsi, A. 2021.
\newblock Computationally instrument-resolution-independent de novo peptide sequencing for high-resolution devices.
\newblock \emph{Nature Machine Intelligence}, 3(5): 420--425.

\bibitem[{Rombach et~al.(2022)Rombach, Blattmann, Lorenz, Esser, and Ommer}]{rombach2022high}
Rombach, R.; Blattmann, A.; Lorenz, D.; Esser, P.; and Ommer, B. 2022.
\newblock High-resolution image synthesis with latent diffusion models.
\newblock In \emph{Proceedings of the IEEE/CVF conference on computer vision and pattern recognition}, 10684--10695.

\bibitem[{Song et~al.(2020)Song, Sohl-Dickstein, Kingma, Kumar, Ermon, and Poole}]{song2020score}
Song, Y.; Sohl-Dickstein, J.; Kingma, D.~P.; Kumar, A.; Ermon, S.; and Poole, B. 2020.
\newblock Score-based generative modeling through stochastic differential equations.
\newblock \emph{arXiv preprint arXiv:2011.13456}.

\bibitem[{Sutskever, Vinyals, and Le(2014)}]{sutskever2014sequence}
Sutskever, I.; Vinyals, O.; and Le, Q.~V. 2014.
\newblock Sequence to sequence learning with neural networks.
\newblock \emph{Advances in neural information processing systems}, 27.

\bibitem[{Tran et~al.(2017)Tran, Zhang, Xin, Shan, and Li}]{tran2017deepnovo}
Tran, N.~H.; Zhang, X.; Xin, L.; Shan, B.; and Li, M. 2017.
\newblock De novo peptide sequencing by deep learning.
\newblock \emph{Proceedings of the National Academy of Sciences}, 114(31): 8247--8252.

\bibitem[{Vaswani et~al.(2017)Vaswani, Shazeer, Parmar, Uszkoreit, Jones, Gomez, Kaiser, and Polosukhin}]{vaswani2017attention}
Vaswani, A.; Shazeer, N.; Parmar, N.; Uszkoreit, J.; Jones, L.; Gomez, A.~N.; Kaiser, {\L}.; and Polosukhin, I. 2017.
\newblock Attention is all you need.
\newblock \emph{Advances in neural information processing systems}, 30.

\bibitem[{Watson et~al.(2023)Watson, Juergens, Bennett, Trippe, Yim, Eisenach, Ahern, Borst, Ragotte, Milles et~al.}]{watson2023novo}
Watson, J.~L.; Juergens, D.; Bennett, N.~R.; Trippe, B.~L.; Yim, J.; Eisenach, H.~E.; Ahern, W.; Borst, A.~J.; Ragotte, R.~J.; Milles, L.~F.; et~al. 2023.
\newblock De novo design of protein structure and function with RFdiffusion.
\newblock \emph{Nature}, 620(7976): 1089--1100.

\bibitem[{Xia et~al.(2024)Xia, Chen, Zhou, Xiaojun, Du, Gao, Tan, Hu, Zheng, and Li}]{xia2024adanovo}
Xia, J.; Chen, S.; Zhou, J.; Xiaojun, S.; Du, W.; Gao, Z.; Tan, C.; Hu, B.; Zheng, J.; and Li, S.~Z. 2024.
\newblock Adanovo: Towards robust$\backslash$emph $\{$De Novo$\}$ peptide sequencing in proteomics against data biases.
\newblock \emph{Advances in Neural Information Processing Systems}, 37: 1811--1828.

\bibitem[{Xu et~al.(2022)Xu, Yu, Song, Shi, Ermon, and Tang}]{xu2022geodiff}
Xu, M.; Yu, L.; Song, Y.; Shi, C.; Ermon, S.; and Tang, J. 2022.
\newblock Geodiff: A geometric diffusion model for molecular conformation generation.
\newblock \emph{arXiv preprint arXiv:2203.02923}.

\bibitem[{Yang et~al.(2024)Yang, Ling, Sun, Liang, Xu, Huang, Xie, He, Li, He et~al.}]{yang2024helixnovo}
Yang, T.; Ling, T.; Sun, B.; Liang, Z.; Xu, F.; Huang, X.; Xie, L.; He, Y.; Li, L.; He, F.; et~al. 2024.
\newblock Introducing $\pi$-HelixNovo for practical large-scale de novo peptide sequencing.
\newblock \emph{Briefings in Bioinformatics}, 25(2): bbae021.

\bibitem[{Yilmaz et~al.(2022)Yilmaz, Fondrie, Bittremieux, Oh, and Noble}]{yilmaz2022casanovo}
Yilmaz, M.; Fondrie, W.; Bittremieux, W.; Oh, S.; and Noble, W.~S. 2022.
\newblock De novo mass spectrometry peptide sequencing with a transformer model.
\newblock In \emph{International Conference on Machine Learning}, 25514--25522. PMLR.

\bibitem[{Zhang, Rao, and Agrawala(2023)}]{zhang2023adding}
Zhang, L.; Rao, A.; and Agrawala, M. 2023.
\newblock Adding conditional control to text-to-image diffusion models.
\newblock In \emph{Proceedings of the IEEE/CVF international conference on computer vision}, 3836--3847.

\bibitem[{Zhang et~al.(2013)Zhang, Fonslow, Shan, Baek, and Yates~III}]{zhang2013protein}
Zhang, Y.; Fonslow, B.~R.; Shan, B.; Baek, M.-C.; and Yates~III, J.~R. 2013.
\newblock Protein analysis by shotgun/bottom-up proteomics.
\newblock \emph{Chemical reviews}, 113(4): 2343--2394.

\bibitem[{Zhou et~al.(2024)Zhou, Chen, Xia, Sizhe~Liu, Ling, Du, Liu, Yin, and Li}]{zhou2024novobench}
Zhou, J.; Chen, S.; Xia, J.; Sizhe~Liu, S.; Ling, T.; Du, W.; Liu, Y.; Yin, J.; and Li, S.~Z. 2024.
\newblock Novobench: Benchmarking deep learning-based$\backslash$emph $\{$De Novo$\}$ sequencing methods in proteomics.
\newblock \emph{Advances in Neural Information Processing Systems}, 37: 104776--104791.

\end{thebibliography}

\end{document}